\documentclass[orivec]{llncs}

\usepackage{amsmath,amssymb,mathabx,amsfonts,wasysym,stmaryrd,cancel}
\usepackage{enumerate}
\usepackage{graphicx}
\graphicspath{{figures/}}
\usepackage{tikz}
\usepackage{listings}
\usepackage{booktabs,tabulary,tabularx,caption,multirow}
\usepackage{bussproofs}
\usepackage{soul,color}
\usepackage{comment,todonotes}
\usepackage{txfonts}

\usepackage{cite}
\usepackage{algorithmic}
\usepackage{textcomp}
\usepackage{bmpsize}
\usepackage{xcolor}
\usepackage{lipsum}
\usepackage[colorlinks=true,urlcolor=black]{hyperref}

\newcommand{\sra}{\mbox{{\footnotesize $\rightarrow$}}}
\newcommand{\logicrule}[2]{\dfrac{{ #1}}{{ #2}}}
\newcommand{\rrule}[2]{{#1} \rightarrow^{\circledast} {#2}}
\newcommand{\ssequent}[3]{[{#1},\ {#2}] \, \mathrel{\vdash_{T}} \, {#3}}
\newcommand{\sequent}[4]{\ssequent{#1}{#2}{\rrule{#3}{#4}}}
\newcommand{\mlf}[2]{{#1} \mathrel{\mid} {#2}}

\newcommand{\cA}[0]{\ensuremath{\cal A}}
\newcommand{\cC}[0]{\ensuremath{\cal C}}
\newcommand{\vars}[1]{\mathit{vars}(#1)}

\begin{document}
\title{Verification of the IBOS Browser Security Properties in Reachability Logic}
\author{Stephen Skeirik\inst{1} \and Jos\'e Meseguer\inst{1} \and Camilo Rocha\inst{2}}
\institute{University of Illinois at Urbana-Champaign\\
	       \email{skeirik2@illinois.edu}\\
	       \email{meseguer@illinois.edu}
	  \and Pontificia Universidad Javeriana Cali\\
           \email{camilo.rocha@javerianacali.edu.co}}
\maketitle

\begin{abstract}
  This paper presents a rewriting logic specification of the Illinois
  Browser Operating System (IBOS) and defines several security
  properties, including the \emph{same-origin policy} (SOP) in
  reachability logic. It shows how these properties can be deductively
  verified using our constructor-based reachability logic theorem
  prover.  This paper also highlights the reasoning techniques used in
  the proof and three modularity principles that have been crucial to
  scale up and complete the verification effort.
\end{abstract}

\section{Introduction}

{\bf Rationale and Origins}.  Web browsers have in fact become
operating systems for a myriad of web-based applications.  Given the
enormous user base and the massive increase in web-based application
areas, browsers have for a long time been a prime target for security
attacks, with a seemingly unending sequence of browser security
violations.  One key reason for this problematic state of affairs is
the enormous size (millions of lines of code) and sheer complexity of
conventional browsers, which make their formal verification a daunting
task.  An early effort to substantially improve browser security by
formal methods was jointly carried out by researchers at Microsoft
Research and the University of Illinois at Urbana-Champaign (UIUC),
who formally specified Internet Explorer (IE) in Maude
\cite{maude-book}, and model checked that formalization finding 13 new
types of unknown address bar or status bar spoofing attacks in it
\cite{IE-maude-analysis}.  
To avoid attacks on those newly found vulnerabilities, they were all corrected in
IE \emph{before} \cite{IE-maude-analysis} was published.  But the
research in \cite{IE-maude-analysis} just uncovered \emph{some} kinds
of possible attacks, and the sheer size and complexity of IE made full
verification unfeasible.  This stimulated a team of systems and formal
methods researchers at UIUC to ask the following question: \emph{could
  formal methods be used from the very beginning in the design of a
  secure browser with a very small trusted code base (TCB) whose
  design could be verified?}  The answer given to this question was
the Maude-based design, model checking verification, and
implementation of the IBOS Browser cum operating system,
\cite{DBLP:conf/osdi/TangMK10,tang-thesis,sasse-thesis,DBLP:conf/facs2/SasseKMT12},
with a 42K line trusted code base (TCB), several orders of magnitude
smaller than the TCBs of commodity browsers.

{\bf Why this Work}.  As further explained in Section
\ref{related-concl}, only a model checking verification of the IBOS
security properties relying on a hand-proof abstraction argument for
its full applicability was possible at the time IBOS was developed
\cite{sasse-thesis,DBLP:conf/facs2/SasseKMT12,rocha-thesis}.
A subsequent attempt at a full deductive verification of IBOS in
\cite{rocha-thesis} had to be abandoned due to the generation of
thousands of proof obligations.  In retrospect, this is not surprising
for two reasons. (1) Many of the symbolic techniques needed to scale
up the IBOS deductive verification effort, including variant
unification and narrowing \cite{variant-JLAP}, order-sorted congruence
closure module axioms \cite{DBLP:conf/fossacs/Meseguer16}, and
variant-based satisfiability
\cite{var-sat-scp,skeirik-meseguer-var-sat-JLAMP}, did not exist at
the time.  In the meantime, those symbolic techniques have been
developed and implemented in Maude. (2) Also missing was a
\emph{program logic} generalizing Hoare logic for Maude specifications
in which properties of concurrent systems specified in Maude could be
specified and verified.  This has been recently addressed with the
development of a \emph{constructor-based reachability logic} for
rewrite theories in 
\cite{DBLP:conf/lopstr/SkeirikSM17,DBLP:journals/fuin/SkeirikSM20}, which extends
prior reachability logic research on verification of conventional
programs using K
in~\cite{DBLP:conf/fm/RosuS12,DBLP:conf/oopsla/RosuS12,DBLP:conf/rta/StefanescuCMMSR14,DBLP:conf/oopsla/StefanescuPYLR16}.
In fact, what has made possible the deductive proof of the IBOS
security properties presented in this paper is precisely the
combination of the strengths from (1) and (2) within the reachability
logic theorem prover that we have developed for carrying out such a
proof.  Implicit in both (1) and (2) are two important proof
obligations. First, both our symbolic reasoning and reachability logic
engines take as input a rewrite theory $\mathcal{R}$.  However, the
correctness of the associated deductions depends on the theory being
\emph{suitable} for symbolic reachability analysis, i.e., its
equations should be ground convergent and sufficiently complete;
therefore, these properties are proof obligations that must be
discharged.  Second, the previous model-checking-based verification
that the IBOS design satisfies certain security properties
\cite{sasse-thesis,DBLP:conf/facs2/SasseKMT12} was based on an
invariant $I_0$.  Our deductive verification uses a slightly different
invariant $I$ that is also \emph{inductive} (as explained in Section
\ref{CBRLOg}).  Thus, we require that $I$ is \emph{at least} as strong
as or stronger than $I_0$ to ensure that our specification of the IBOS
security properties does not miss any cases covered by the prior
work. Both of these important proof obligations have been fully
checked as explained in \cite{skeirik-thesis}.  Last, but not least,
as we further explain in Section \ref{related-concl}, the IBOS browser
security goals remain as relevant and promising today as when IBOS was
first developed, and this work bring us closer to achieving those
goals.

{\bf Main Contributions}. They include:

\begin{itemize}
  \item The first full \emph{deductive verification of the IBOS
    browser} as explained above.
  \item A general \emph{modular proof methodology} for scaling up
    reachability logic proofs of object-based distributed systems that
    has been invaluable for verifying IBOS, but has a much wider
    applicability to general distributed system verification.
  \item A substantial and useful \emph{case study} that can be of help
    to other researchers interested in both browser verification and
    distributed system verification.
  \item New capabilities of the \emph{reachability logic prover},
    which in the course of this research has evolved from the original
    prototype reported in \cite{DBLP:conf/lopstr/SkeirikSM17} to a
    first prover version to be released in the near future.
\end{itemize}

{\bf Plan of the Paper}.  Preliminaries are gathered in Section
\ref{prelims}.  Reachability Logic and invariant verification are
presented in Section \ref{CBRLOg}.  IBOS, its rewriting logic Maude
specification, and the specification of its security properties are
explained in Section \ref{IBOS-Sect}.  The deductive proof of those
IBOS properties and the modular proof methodology used are described
in Section \ref{IBOS-Proof}.  Section \ref{related-concl} discusses
related work and concludes the paper.

\section{Preliminaries on Equational and Rewriting Logic}
\label{prelims}

We present some preliminaries on order-sorted equational logic and
rewriting logic. The material is adapted from
\cite{osa1,tarquinia,20-years}.

{\bf Order-Sorted Equational Logic}.  We assume the basic notions of
order-sorted (abbreviated OS) signature $\Sigma$, $\Sigma$-term $t$,
$\Sigma$-algebra $A$, and $\Sigma$-homomorphism $f:A \rightarrow B$
\cite{osa1,tarquinia}.  Intuitively, $\Sigma$ defines a partially
ordered set of sorts $(S,\leq)$, which are interpreted in a
$\Sigma$-algebra $A$ with carrier family of sets $A=\{A_{s}\}_{s \in
  S}$ as sort containments.  For example, if we have a sort inclusion
$\mathit{Nat} < \mathit{Int}$, then we must have $A_{\mathit{Nat}}
\subseteq A_{\mathit{Int}}$.  An operator, say $+$, in $\Sigma$ may
have several related typings, e.g., $+: \mathit{Nat} \; \mathit{Nat}
\rightarrow \mathit{Nat}$ and $+: \mathit{Int} \; \mathit{Int}
\rightarrow \mathit{Int}$, whose interpretations in an algebra $A$
must agree when restricted to subsorts.  The OS algebras over
signature $\Sigma$ and their homomorphisms form a category
$\textbf{OSAlg}_{\Sigma}$.  Furthermore, under mild syntactic
conditions on $\Sigma$, the term algebra $T_{\Sigma}$ is initial
\cite{tarquinia}; all signatures are assumed to satisfy these
conditions.

An $S$-sorted set $X=\{X_{s}\}_{s \in S}$ of \emph{variables},
satisfies $s \not= s' \Rightarrow X_{s}\cap X_{s'}=\emptyset$, and the
variables in $X$ are always assumed disjoint from all constants in
$\Sigma$.  The $\Sigma$-\emph{term algebra} on variables $X$,
$T_{\Sigma}(X)$, is the \emph{initial algebra} for the signature
$\Sigma(X)$ obtained by adding to $\Sigma$ the variables $X$ \emph{as
  extra constants}.  Given a $\Sigma$-algebra $A$, an
\emph{assignment} $a$ is an $S$-sorted function $a \in [X \rightarrow
  A]$ mapping each variable $x \in X_{s}$ to a value $a(x)\in A_{s}$
for each $s\in S$.  Each such assignment uniquely extends to a
$\Sigma$-homomorphism $\_a : T_{\Sigma}(X) \rightarrow A$, so that if
$x \in X_{s}$, then $x\, a = a(x)$.  In particular, for
$A=T_{\Sigma}(X)$, an assignment $\sigma \in [X \rightarrow
  T_{\Sigma}(X)]$ is called a \emph{substitution} and uniquely extends
to a $\Sigma$-homomorphism $\_\sigma : T_{\Sigma}(X) \rightarrow
T_{\Sigma}(X)$.  Define $\mathit{dom}(\sigma)=\{x \in X \mid x \not= x
\sigma\}$ and $\mathit{ran}(\sigma)= \bigcup_{x \in
  \mathit{dom}(\sigma)} \mathit{vars}(x \sigma)$.

We assume familiarity with the language of first-order logic with
equality.  In particular, given a $\Sigma$-formula $\varphi$, we
assume familiarity with the satisfaction relation $A ,a \models
\varphi$ for a $\Sigma$-algebra $A$ and assignment $a \in
       [\mathit{fvars}(\varphi) \sra A]$ for the \textit{free variables}
       $\mathit{fvars}(\varphi)$ of $\varphi$.  Then, $\varphi$ is
       \emph{valid} in $A$, denoted $A \models \varphi$, iff $\forall
       \; a \in [\mathit{fvars}(\varphi) \sra A]$ $A ,a \models
       \varphi$, and is \emph{satisfiable} in $A$ iff $\exists \; a
       \in [\mathit{fvars}(\varphi) \sra A]$ $A ,a \models \varphi$.
       Let $\mathit{Form}(\Sigma)$ (resp. $\mathit{QFForm}(\Sigma)$)
       denote the set of $\Sigma$-formulas (resp. quantifier free
       $\Sigma$-formulas).

An OS \emph{equational theory} is a pair $T=(\Sigma,E)$, with $E$ a
set of (possibly conditional) $\Sigma$-equations.
$\textbf{OSAlg}_{(\Sigma,E)}$ denotes the full subcategory of
$\textbf{OSAlg}_{\Sigma}$ with objects those $A \in
\textbf{OSAlg}_{\Sigma}$ such that $A \models E$, called the
$(\Sigma,E)$-\emph{algebras}.  The inference system in
\cite{tarquinia} is \emph{sound and complete} for OS equational
deduction.  $E$-\emph{equality}, i.e., provability $E \vdash u=v$, is
written $u =_{E} v$.  $\textbf{OSAlg}_{(\Sigma,E)}$ has an
\emph{initial algebra} $T_{\Sigma/E}$ \cite{tarquinia}.  Given a
system of $\Sigma$ equations $\phi=u_{1}=v_{1} \wedge \ldots \wedge
u_{n}=v_{n}$, an $E$-\emph{unifier} for $\phi$ is a substitution
$\sigma$ such that $u_{i} \sigma=_{E} v_{i}\sigma$, $1 \leq i \leq n$;
an $E$-\emph{unification algorithm} for $(\Sigma,E)$ generates a
\emph{complete set} of $E$-unifiers $\mathit{Unif}_{\!E}(\phi)$ for
any system $\phi$ in the sense that, up to $E$-equality, any
$E$-unifier $\sigma$ of $\phi$ is a substitution instance of some
unifier $\theta \in \mathit{Unif}_{\!E}(\phi)$.

{\bf Rewriting Logic}.  A \emph{rewrite theory} $\mathcal{R}=(\Sigma,E
\cup B,R)$, with $(\Sigma,E \cup B)$ an OS-equational theory with
equations $E$ and structural axioms $B$ (typically any combination of
associativity, commutativity, and identity), and $R$ a collection of
rewrite rules, specifies a \emph{concurrent system} whose states are
elements of the initial algebra $T_{\Sigma/E \cup B}$ and whose
\emph{concurrent transitions} are specified by the rewrite rules $R$.
The concurrent system thus specified is the \emph{initial reachability
  model} $\mathcal{T}_{\mathcal{R}}$ associated to $\mathcal{R}$
\cite{bruni-meseguer-tcs,20-years}.

Maude \cite{maude-book} is a declarative programming language whose
programs are exactly rewrite theories.  To be executable in Maude, a
rewrite theory $\mathcal{R}=(\Sigma,E \cup B,R)$ should satisfy some
\emph{executability conditions} spelled out below.  Recall the
notation for term positions, subterms, and replacement from
\cite{dershowitz-jouannaud}: (i) positions in a term are marked by
strings $p \in \mathbb{N}^{*}$ specifying a path from the root, (ii)
$t|_{p}$ denotes the subterm of term $t$ at position $p$, and (iii)
$t[u]_{p}$ denotes the result of \emph{replacing} subterm $t|_{p}$ at
position $p$ by $u$.

\begin{definition} \label{rewrite-theory-defns}
	An \emph{executable rewrite theory} is a 3-tuple
  $\mathcal{R}=(\Sigma,E \cup B,R)$ with $(\Sigma,E \cup B)$ an OS
  equational theory with $E$ possibly conditional and $R$ a set of
  possibly conditional $\Sigma$-\emph{rewrite rules}, i.e., sequents
  $l \rightarrow r \; \mathit{if}\; \phi$, with $l,r \in
  T_{\Sigma}(X)_{s}$ for some $s \in S$, and $\phi$ a quantifier-free
  $\Sigma$-formula.  We further assume that:
	\begin{enumerate}
		\item $B$ is a collection of associativity and/or
        commutativity and/or identity axioms and
        $\Sigma$ is $B$-\emph{preregular} \cite{maude-book}.

		\item Equations $E$, oriented as
		rewrite rules $\vec{E}$, are \emph{convergent} modulo $B$
		\cite{DBLP:journals/jlp/LucasM16}.

		\item Rules $R$ are \emph{ground coherent} with the equations
		$E$ modulo $B$ \cite{DBLP:journals/jlp/DuranM12}.
	\end{enumerate}
	The \emph{one-step} $R,B$-\emph{rewrite relation} $t
  \rightarrow_{R,B} t'$ holds iff there is a rule $l\rightarrow r \;
  \mathit{if}\; \phi \in R$, a ground substitution $\sigma \in [Y \sra
    T_{\Sigma}]$ with $Y$ the rule's variables, and position $p$ where
  $t|_{p}=_{B} l \sigma$, $t'=t[r \sigma]_{p}$, and $T_{\Sigma/E \cup
    B} \models \phi \sigma$.  Let $\rightarrow^*_{R,B}$ denote the
  reflexive-transitive closure of the rewrite relation
  $\rightarrow_{R,B}$.
\end{definition}

Intuitively, conditions (1)--(2) ensure that the initial algebra
$T_{\Sigma/E \cup B}$ is isomorphic to the \emph{canonical term
  algebra} $C_{\Sigma/E,B}$, whose elements are $B$-equivalence
classes of $\vec{E},B$-\emph{canonical} ground $\Sigma$-terms, where
$v$ is the $\vec{E},B$-\emph{canonical form} of a term $t$, denoted
$u=t!_{\vec{E},B}$, iff: (i) $t \rightarrow^{*}_{\vec{E},B} u$, and
(ii) $(\nexists v\in T_\Sigma)\; u \rightarrow_{\vec{E},B} v$.  By
$\vec{E}$ convergent modulo $B$, $t!_{\vec{E},B}$ is unique up to
$B$-equality \cite{DBLP:journals/jlp/LucasM16}.  Adding (3) ensures
that ``computing $\vec{E},B$-canonical forms before performing
$R,B$-rewriting'' is a \emph{complete} strategy for rewriting with the
rules $R$ module equations $E$.  That is, if $t \rightarrow_{R,B} t'$
and $t!_{\vec{E},B} = u$, then there exists a $u'$ such that $u
\rightarrow_{R,B} u'$ and $t'!_{\vec{E},B} =_{B} u'!_{\vec{E},B}$.  We
refer to
\cite{20-years,DBLP:journals/jlp/LucasM16,DBLP:journals/jlp/DuranM12}
for more details.

Conditions (1)--(3) allow a simple and intuitive description of the
\emph{initial reachability model} $\mathcal{T}_{\mathcal{R}}$
\cite{bruni-meseguer-tcs} of $\mathcal{R}$ as the \emph{canonical
  reachability model} $\mathcal{C}_{\mathcal{R}}$ whose states are the
elements of the \emph{canonical term algebra} $C_{\Sigma/E,B}$, and
where the one-step transition relation $[u] \rightarrow_{\mathcal{R}}
[v]$ holds iff $u \rightarrow_{R,B} u'$ and $[u'!_{\vec{E},B}] = [v]$.
Finally, if $u \rightarrow_{R,B} u'$ via rule $(l\rightarrow r \;
\mathit{if}\; \phi) \in R$ and a ground substitution $\sigma \in [Y
  \sra T_{\Sigma}]$, then checking if condition $T_{\Sigma/E \cup B}
\models \phi \sigma$ holds is \emph{decidable} by reducing terms in
$\phi \sigma$ to $\vec{E},B$-canonical form.

An OS-subsignature $\Omega \subseteq \Sigma$ is called a
\emph{constructor subsignature} for an OS equational theory $(\Sigma,E
\cup B)$ where $\vec{E}$ is convergent modulo $B$ iff $\forall t\in
T_{\Sigma}\ t!_{\vec{E},B}\in T_{\Omega}$.  Furthermore, the
constructors $\Omega$ are then called \emph{free} modulo axioms
$B_{\Omega} \subseteq B$ iff, as $S$-sorted sets,
$C_{\Sigma/E,B}=T_{\Omega/B_{\Omega}}$.  This assumption gives a
particularly simple description of the states of the canonical
reachability model $\mathcal{C}_{\mathcal{R}}$ as
$B_{\Omega}$-equivalence classes of ground $\Omega$-terms.  As
explained in Section \ref{CBRLOg}, this simple
\emph{constructor-based} description is systematically exploited in
reachability logic.

An executable rewrite theory $\mathcal{R}=(\Sigma,E \cup B,R)$ with
constructor subsignature $\Omega$ is called \emph{topmost} iff the
poset of sorts $(S,\leq)$ has a maximal element, call it
$\mathit{State}$, such that: (i) for all rules $(l \rightarrow r \;
\mathit{if}\; \phi)\in R$, $l,r \in T_{\Omega}(X)_{\mathit{State}}$;
and (ii) for any $f: s_{1} \ldots s_{n} \rightarrow s$ in $\Omega$
with $s \leq \mathit{State}$ we have $s_{i} \not\leq \mathit{State}$,
$1 \leq i \leq n$.  This ensures that if $[u] \in
\mathcal{C}_{\mathcal{R}}=T_{\Omega/B_{\Omega}}$, and $[u] \in
\mathcal{C}_{\mathcal{R}, \mathit{State}}$, then all rewrites $u
\rightarrow_{R,B} u'$ happen at the top position $\epsilon$.  This
topmost requirement is easy to achieve in practice.  In particular, it
can \emph{always} be achieved for \emph{object-based rewrite
  theories}, which we explain next.

{\bf Object-Based Rewrite Theories}.  Most distributed systems,
including the IBOS browser, can be naturally modeled by
\emph{object-based rewrite theories}.  We give here a brief
introduction and refer to \cite{ooconc,maude-book} for more details.
The \emph{distributed state} of an object-based system, called a
\emph{configuration}, is modeled as a \emph{multiset} or ``soup'' of
objects and messages built up by an \emph{associative-commutative}
binary multiset union operator (with juxtaposition syntax) $\_ \, \_ :
\mathit{Conf} \; \mathit{Conf} \rightarrow \mathit{Conf}$ with
identity $\mathit{null}$.  The sort $\mathit{Conf}$ has two subsorts:
a sort $\mathit{Object}$ of \emph{objects} and a sort $\mathit{Msg}$
of \emph{messages} ``traveling'' in the configuration from a sender
object to a receiver object.  The syntax for messages is
user-definable, but it is convenient to adopt a conventional syntax
for objects as record-like structures of the form: $\langle \, o \mid
a_{1}(v_{1}),\ldots, a_{n}(v_{n})\rangle$, where $o$ is the object's
name or \emph{object identifier}, belonging to a subsort of a general
sort $\mathit{Oid}$, and $a_{1}(v_{1}),\ldots, a_{n}(v_{n})$ is a
\emph{set} of object \emph{atributes} of sort $\mathit{Att}$ built
with an associative-commutative union operator $\_ \; , \_ :
\mathit{Atts} \; \mathit{Atts} \rightarrow \mathit{Atts}$, with
$\mathit{null}$ as identity element and with $\mathit{Att} <
\mathit{Atts}$.  Each $a_{i}$ is a constructor operator $a_{i}: s_{i}
\rightarrow \mathit{Att}$ so that the \emph{data value} $v_{i}$ has
sort $s_{i}$.  Objects can be classified in \emph{object classes}, so
that a class $C$ has an associated subsort $\mathit{C.Oid} <
\mathit{Oid}$ for its object identifiers and associated attribute
constructors $a_{i}: s_{i} \rightarrow \mathit{Att}$, $1 \leq i \leq
n$.  Usually, a configuration may have many objects of the same class,
each with a different object identifier; but some classes (e.g., the
$\mathit{Kernel}$ class in IBOS) are \emph{singleton classes}, so that
only one object of that class, with a fixed name, will apear in a
configuration.  Another example in the IBOS specification is the
singleton class $\mathit{Display}$.  The single display object
represents the rendering of the web page shown to the user and has the
form: $< \textit{display} \mid \textit{displayContent}(D),
\textit{activeTab}(\textit{WA}) >$, where the $\textit{activeTab}$
attribute constructor contains a reference to the web process that the
user has selected (each tab corresponds to a different web process)
and the $\textit{displayContent}$ constructor encapsulates the web
page content currently shown on the display.  Not all configurations
of objects and messages are sensible.
A configuration is \emph{well-formed} iff it satisfies the following
two requirements: (i) \emph{unique object identifiers}: each object
has a unique name different from all other object names; and (ii)
\emph{uniqueness of object attributes}: within an object, each object
attribute \emph{appears only once}; for example, an object like $<
\textit{display} \mid \textit{displayContent}(D),
\textit{activeTab}(\textit{WA}) , \textit{activeTab}(\textit{WA'}) >$
is nonsensical.

The \emph{rewrite rules} $R$ of an object-based rewrite theory
$\mathcal{R}=(\Sigma,E \cup B,R)$ have the general form $l\rightarrow
r \; \mathit{if}\; \phi$, where $l$ and $r$ are terms of sort
$\mathit{Conf}$.  Intuitively, $l$ is a pattern describing a
\emph{local fragment} of the overall configuration, so that a
substitution instance $l \sigma$ describing a concrete such fragment
(e.g., two objects, or an object and a message) can be rewritten to a
new subfragment $r \sigma$, provided the rule's condition $\phi
\sigma$ holds (see Section \ref{IBOS-SSpec} for an example rule).
Classes can be structured in \emph{multiple inheritance hierarchies},
where a subclass $C'$ of class $C$, denoted $C' < C$, may have
additional attributes and has a subsort $\mathit{C'.Oid} <
\mathit{C.Oid}$ for its object identifiers.  By using extra variables
$\mathit{Attrs_{j}}$ of sort $\mathit{Atts}$ for ``any extra
attributes'' that may appear in a subclass of any object $o_{j}$ in
the left-hand side patterns $l$ of a rule, rewrite rules can be
\emph{automatically inherited by subclasses} \cite{ooconc}.
Furthermore, a subclass $C' < C$ may have \emph{extra rules}, which
may modify both its superclass attributes and its additional
attributes.

An object-based rewrite theory $\mathcal{R}=(\Sigma,E \cup B,R)$
\emph{can easily be made topmost} as follows: (i) we add a fresh now
sort $\mathit{State}$ and an ``encapsulation operator,'' say, $\{\_\}
: \mathit{Conf} \rightarrow \mathit{State}$, and (ii) we trasform each
rule $l\rightarrow r \; \mathit{if}\; \phi$ into the rule $\{l \;
C\}\rightarrow \{r \; C\} \; \mathit{if}\; \phi$, where $C$ is a fresh
variable of sort $\mathit{Conf}$ modeling ``the rest of the
configuration,'' which could be empty by the identity axiom for
$\mathit{null}$.

\section{Constructor-Based Reachability Logic} \label{CBRLOg}

Constructor-based reachability logic
\cite{DBLP:conf/lopstr/SkeirikSM17,DBLP:journals/fuin/SkeirikSM20} 
is a partial correctness logic
generalizing Hoare logic in the following sense.  In Hoare logic we
have state predicates $A,B,C,\ldots$ and formulas are Hoare triples
$\{A\} \, p \, \{B\}$ where $A$ is the \emph{precondition}, $B$ is the
\emph{postcondition}, and $p$ is the \emph{program} we are reasoning
about.  Since a Maude program is a rewrite theory $\mathcal{R}$, a
Hoare triple in Maude has the form $\{A\} \, \mathcal{R} \, \{B\}$,
with the expected \emph{partial correctness semantics}.  But we can be
more general and consider \emph{reachability logic formulas} of the
form: $A \rightarrow^{\circledast} B$, with $A$ the
\emph{precondition} (as in Hoare logic) but with $B$ what we call the
formula's \emph{midcondition}.  That is, $B$ need not hold \emph{at
  the end} of a terminating computation, as in Hoare logic, but just
\emph{somewhere in the middle} of such a computation.  A topmost
rewrite theory $\mathcal{R}$ \emph{satisfies} $A
\rightarrow^{\circledast} B$, written $\mathcal{R} \models A
\rightarrow^{\circledast} B$, iff along any terminating computation
from an initial state $[u]$ satisfying $A$ there is some intermediate
state satisfying $B$.  More precisely:

\begin{definition} \label{reach-logic-form-defn}
	Let $\mathcal{R}=(\Sigma,E\cup B,R)$ be topmost with top sort
  $\mathit{State}$ and with free constructors $\Omega$ modulo
  $B_\Omega$, and let $A$ and $B$ be state predicates for states of
  sort $\mathit{State}$, so that $\llbracket A \rrbracket$ and
  $\llbracket B \rrbracket$ denote the respective subsets of
  $T_{\Omega/B_{\Omega}, \mathit{State}}$ defined by $A$ and
  $B$. Furthermore, let $T$ be a state predicate of \emph{terminating
    states} so that $\llbracket T \rrbracket \subseteq
  \mathit{Term}_{\mathcal{R}}$, where $\mathit{Term}_{\mathcal{R}} =
  \{ [u] \in T_{\Omega/B_{\Omega},\mathit{State}} \mid
  (\not\exists[v])\; [u] \rightarrow_{\mathcal{R}} [v]\}$.  Then, a
  reachability formula $A \rightarrow^{\circledast} B$ holds for the
  canonical reachability model $\mathcal{C}_{\mathcal{R}}$ of
  $\mathcal{R}$ relative to $T$ in the \emph{all paths satisfaction
    relation}, denoted $\mathcal{R} \models_{T}^{\forall} A
  \rightarrow^{\circledast} B$, iff for every sequence $[u_{0}]
  \rightarrow_{\mathcal{R}}[u_{1}] \ldots
             [u_{n-1}]\rightarrow_{\mathcal{R}}[u_{n}]$ with $[u_{0}]
             \in \llbracket A \rrbracket$ and $[u_{n}]\in \llbracket T
             \rrbracket$ there exists $k$, $0 \leq k \leq n$ such that
             $[u_{k}] \in \llbracket B \rrbracket$.

When $\llbracket T \rrbracket = \mathit{Term}_{\mathcal{R}}$, we
abbreviate the relation $\models_{T}^{\forall}$ to just $\models$.
\end{definition}

We can define the satisfaction of a Hoare triple $\{A\}\, \mathcal{R}
\,\{B\}$ as syntactic sugar for the satisfaction relation $\mathcal{R}
\models A \rightarrow^{\circledast} (B\wedge
\mathit{Term}_{\mathcal{R}})$.  As explained in
\cite{DBLP:journals/fuin/SkeirikSM20}, the case of a Hoare logic for an
\emph{imperative} programming language $\mathcal{L}$ is obtained as
syntactic sugar for a special class of reachability logic formulas for
a rewrite theory $\mathcal{R}_{\mathcal{L}}$ giving a rewriting logic
semantics to the programming language $\mathcal{L}$.

But in what sense is such a reachability logic
\emph{constructor-based}?  In the precise sense that the state
predicates $A,B,C,\ldots$ used in the logic are \emph{constrained
  constructor pattern predicates} that exploit for symbolic purposes
the extreme simplicity of the constructor theory
$(\Omega,B_{\Omega})$, which is much simpler than the equational
theory $(\Sigma,E\cup B)$.

\begin{definition} \label{ctor-pat-pred-defn}
	Let $(\Sigma,E \cup B)$ be convergent modulo $B$ and sufficiently
  complete with respect to $(\Omega,B_{\Omega})$, i.e.,
  $C_{\Sigma/E,B}=T_{\Omega/B_{\Omega}}$.  A \emph{constrained
    constructor pattern} is an expression $(u\mid \phi)$ such that
  $(u,\phi) \in T_\Omega(X)\!\times\! \mathit{QFForm}(\Sigma)$.  The
  set of \emph{constrained constructor pattern \text{predicates}}
  $\mathit{PatPred}(\Omega,\Sigma)$ is defined inductively over
  $T_\Omega(X)\!\times\! \mathit{QFForm}(\Sigma)$ by adding $\bot$ and
  closing under ($\vee\!$) and ($\wedge\!$).  We let capital letters
  $A,B,\ldots,P,Q,\ldots$ range over
  $\mathit{PatPred}(\Omega,\Sigma)$.  The \emph{semantics} of $A\in
  \mathit{PatPred}(\Omega,\Sigma)$ is the subset $\llbracket A
  \rrbracket \subseteq C_{\Sigma/E,B}$ such that:
	\begin{enumerate}[(i)]
		\item $\llbracket \bot \rrbracket = \emptyset$,
		\item $\llbracket u \mid \varphi \rrbracket = \{[(u \rho)!]_{B_{\Omega}} \in
		C_{\Sigma/E,B} \mid \rho \in  [X \sra T_{\Omega}] \, \wedge \,   C_{\Sigma/E,B}   \models \varphi \rho\}$,
		\item $\llbracket A \vee B \rrbracket$ = $\llbracket A \rrbracket \cup \llbracket B \rrbracket$, and
		\item $\llbracket A \wedge B \rrbracket$ = $\llbracket A \rrbracket \cap \llbracket B \rrbracket$.
	\end{enumerate}
For any sort $s$, let $\mathit{PatPred}(\Omega,\Sigma)_s\subseteq
\mathit{PatPred}(\Omega,\Sigma)$ where $A\in
\mathit{PatPred}(\Omega,\Sigma)_s$ iff each subpattern $(u\mid \phi)$
of $A$ has $u\in T_\Omega(X)_s$.
	$A\in \mathit{PatPred}(\Omega,\Sigma)$ is \emph{normal} if it has no subpattern of the form $P \wedge Q$.
	If $B_\Omega$ has a finitary unification algorithm, normalization is effectively
	computable by (disjoint) $B_\Omega$ unification
\cite{DBLP:conf/lopstr/SkeirikSM17,DBLP:journals/fuin/SkeirikSM20}.
\end{definition}

We can now fully specify our constructor-based reachability logic: it
is a reachability logic for topmost rewrite theories
$\mathcal{R}=(\Sigma,E\cup B,R)$ with top sort $\mathit{State}$ and
with free constructors $\Omega$ modulo $B_\Omega$ whose set of state
predicate formulas is
$\mathit{PatPred}(\Omega,\Sigma)_{\mathit{State}}$.

\textbf{Reachability Logic Proof Rules
  \cite{DBLP:conf/lopstr/SkeirikSM17,DBLP:journals/fuin/SkeirikSM20}.} We review our proof system for
reachability logic that was proved sound with respect to the all-paths
satisfaction relation in \cite{DBLP:conf/lopstr/SkeirikSM17,DBLP:journals/fuin/SkeirikSM20}.  The
proof rules derive sequents of the form
$\ssequent{\cA}{\cC}{\rrule{A}{B}}$, where $\cA$ and $\cC$ are finite
sets of reachability formulas, $T\subseteq
\mathit{Term}_{\mathcal{R}}$, and reachability formula $\rrule{A}{B}$
is normalized.  Formulas in $\cA$ are called {\em axioms}; those in
$\cC$ are called {\em circularities}. The proof system has three
rules: \textsc{Step}, \textsc{Axiom}, and \textsc{Subsumption} as well
as derived rules \textsc{Case}, \textsc{Split}, and
\textsc{Substitution}. See the brief overview in Table
\ref{RulesOverview}.  The derived rules are explained in Appendix
\ref{der-pr-rules}.

{\captionsetup{justification=raggedright,labelsep=none,labelformat=empty,singlelinecheck=off}
\begin{table*}
	\raggedright
	\captionof{table}{\hspace{36pt} \textbf{\ul{Table 1: Overview of Proof Rules}}}\label{RulesOverview}

	\vspace{5pt}
	\hspace{36pt} \textit{Assumptions:}\\
	\hspace{36pt} 1. $\mathcal{R}=(\Sigma,E\!\cup\! B,R)$ is suff. comp. w.r.t. $(\Omega,B_\Omega) \text{ with } R=\{l_i\rightarrow r_i \text{ if } \phi_i\}_{i\in I}$\\
	\hspace{36pt} 2. $\rrule{P}{(\bigvee\nolimits_j v_j\mid \psi_j)}\in\cA$
	\vspace{5pt}

	\centering
	\begin{tabular}{@{}lcl@{}}
	\toprule
	Name & Rule & Condition\\
	\midrule
	\textsc{Step}$^\text{*}$  &
	$\logicrule
	{\bigwedge_{i\in I,\alpha\in Y(i)}
		\sequent{\cA \cup \cC}{\emptyset}
		{(\mlf{r_i}{\varphi'\!\! \wedge\! \phi_i}) \alpha}
		{\!B \alpha}}
	{\sequent{\cA}{\cC}{\mlf{u}{\varphi}}{B}}$\vspace*{10pt}

	& \\

	\textsc{Axiom}\ \ &
	$\logicrule
	{\bigwedge\nolimits_j
		\sequent
		{\cA}
		{\cC}{(\mlf{v_j \alpha}{\varphi \wedge \psi_j \alpha})}{B}}
	{\sequent
		{ \cA}
		{\cC}{(\mlf{u}{\varphi})}{B}}$\vspace*{10pt}

	& $\llbracket u\!\mid\!\varphi \rrbracket\subseteq \llbracket P\alpha \rrbracket$\\

	\textsc{Sub.} &
	$\logicrule
	{}
	{\sequent{\cA}{\cC}{A}{B}}$
	& $\llbracket A \rrbracket\subseteq \llbracket B \rrbracket$\vspace*{3pt} \\

	\bottomrule
	\end{tabular}\phantom{a}\smallskip\\

	\raggedright
	\hspace{36pt} $^\text{*} \varphi' = \varphi \wedge \bigwedge \{ \neg \psi\beta\ \mid\, (w\!\mid\! \psi)\in B \wedge \exists \beta\ w\beta=u \}$,\ \
	$Y(i)=\mathit{Unif}_{\! B_\Omega}\!(u,l_i)$\\
\end{table*}
}

This inference system has been mechanized using the Maude rewriting
engine
\cite{DBLP:conf/lopstr/SkeirikSM17,DBLP:journals/fuin/SkeirikSM20}.  
Let us say a few words
about its \emph{automation}.  The \textsc{Step} rule can be automated
by \emph{narrowing}, i.e., symbolic rewriting where the lefthand side
of a rewrite rule $l\rightarrow r \text{ if } \phi$, instead of being
\emph{matched} by a term $t$ to be rewritten, only
$B_\Omega$-\emph{unifies} with it.  Assuming $B_\Omega$ has a finitary
unification algorithm, the narrowing relation lifted to constrained
constructor patterns is \emph{decidable}. The \textsc{Axiom} and
\textsc{Subsumption} rules require automating a constrained
constructor pattern subsumption check of the form $\llbracket u\mid
\varphi\rrbracket \subseteq \llbracket v\mid\phi\rrbracket$; this can
be shown to hold by $B_{\Omega}$-matching $v\alpha=_{B_{\Omega}}u$ and
checking whether $T_{\Sigma/E \cup B} \models \varphi \Rightarrow
\phi\alpha$. Since $B_\Omega$-matching is decidable, the only
undecidable check is the implication's validity. Our tool employs
multiple heuristics to check whether $T_{\Sigma/E \cup B} \models
\varphi \Rightarrow \phi\alpha$. Given a goal
$G=\sequent{\cA}{\cC}{A}{B}$, there are two final operations needed:
(i) checking whether the proof failed, i.e., whether $\llbracket
A\rrbracket \cap \llbracket T\rrbracket\neq \emptyset$ but not
$\llbracket A \rrbracket \subseteq \llbracket B \rrbracket$, (ii) as
an optimization, discarding a goal $G$ where $A$'s constraint is
unsatisfiable. These two checks are also undecidable in general, so we
rely on best-effort heuristics. Interestingly, thanks to their use of
the symbolic techniques for variant unification and satisfiability
\cite{variant-JLAP,var-sat-scp,skeirik-meseguer-var-sat-JLAMP}, and
for order-sorted congruence closure module axioms
\cite{DBLP:conf/fossacs/Meseguer16} mentioned in the Introduction,
these heuristics were sufficient to complete all reachability logic
proofs of the IBOS properties without relying on an external prover.

\textbf{Proving Inductive Invariants
  \cite{DBLP:journals/fuin/SkeirikSM20,GRT-coh-compl-symb-meth-TR}.}
For a transition system $\mathcal{Q} =(Q,\rightarrow_{\mathcal{Q}})$
and a subset $Q_{0} \subseteq Q$ of \emph{initial states}, a subset $I
\subseteq Q$ is called an \emph{invariant} for $\mathcal{Q}$ from
$Q_{0}$ iff for each $a \in Q_{0}$ and $b \in Q$, $a
\rightarrow^{*}_{\mathcal{Q}} b$ implies $b \in I$, where
$\rightarrow^{*}_{\mathcal{Q}}$ denotes the reflexive-transitive
closure of $\rightarrow_{\mathcal{Q}}$.
A subset $A \subseteq Q$ is called \emph{stable} in $\mathcal{Q}$ iff
for each $a \in A$ and $b \in Q$, $a \rightarrow_{\mathcal{Q}} b$
implies $b \in A$.  An invariant $I$ for $\mathcal{Q}$ from $Q_{0}$ is
called \emph{inductive} iff $I$ is stable.
We instantiate this generic framework to prove invariants over a
topmost rewrite theory $\mathcal{R}=(\Sigma,E\cup B,R)$ with
$(\Sigma,B,\vec{E})$ convergent and sufficiently complete with respect
to $(\Omega,B_\Omega)$ and consider the transition system induced by
$\mathcal{R}$ over sort $\mathit{State}$, i.e., $(C_{\Sigma/E\cup
  B,\mathit{State}},\rightarrow_\mathcal{R})$.

To prove an invariant $I$ from $Q_{0}$ over $\mathcal{R}$, we use a
simple theory transformation mapping topmost theory $\mathcal{R}$ to a
theory $\mathcal{R}_{\mathit{stop}}$ having a fresh operator $[\_]$
and a rule $\mathit{stop} : c(\vec{x}) \rightarrow [c(\vec{x})]$ for
each constructor $c$ of sort $\mathit{State}$.  Then, by Corollary 1
in \cite{DBLP:conf/lopstr/SkeirikSM17}, to prove $I$ is an invariant
from $Q_{0}$ over $\mathcal{R}$ we prove $Q_{0} \subseteq I$ and that
the reachability formula $\rrule{I\sigma}{[I]}$ holds over
$\mathcal{R}$, where: (i) $I\sigma$ is a renaming of $I$ with
$\vars{I}\cap\vars{I\sigma}=\emptyset$ and (ii) if $I=(u \mid\varphi)$
then $[I]=([u]\mid\varphi)$. If $I$ is inductive and $I=u\mid\varphi$,
the proof of $\rrule{I\sigma}{[I]}$ proceeds as follows:

\begin{enumerate}
	\item  The initial sequent is $[\emptyset,\rrule{I\sigma}{[I]}] \vdash_{[]} \rrule{I\sigma}{[I]}$.
	\item  Apply the \textsc{Step} rule; based on which rule $\zeta : l\rightarrow r \text{ if } \phi$ was used, obtain:
	\begin{enumerate}
		\item if $\zeta$ is \textit{stop}, $[\rrule{I\sigma}{[I]},\emptyset] \vdash_{[]} \rrule{[I\sigma]}{[I]}$;
		\item otherwise, $\bigwedge_{\alpha\in\mathit{Unif}_{\!B_{\!\Omega}}\!(u,l)} [\rrule{I\sigma}{[I]},\emptyset] \vdash_{[]} \rrule{(r\mid \varphi\wedge\phi)\alpha}{[I]}$.
	\end{enumerate}
	\item For case (2a), apply \textsc{Subsumption}. For case
          (2b), apply zero or more derived rules to obtain sequents of
          the form:
	$[\rrule{I\sigma}{[I]},\emptyset] \vdash_{[]} \rrule{A}{[I]}$.
	\item Since $I$ is assumed inductive, we have
       $\llbracket A \rrbracket \subseteq \llbracket I\sigma
       \rrbracket$.
      Thus, apply \textsc{Axiom} to derive
     $[\rrule{I\sigma}{[I]},\emptyset] \vdash_{[]} \rrule{[I]}{[I]}$.
	\item Finally, apply \textsc{Subsumption}.
\end{enumerate}

Since all the IBOS security properties \emph{are invariants}, all our
proofs follow steps (1)--(5) above.

\section{IBOS and its Security Properties} \label{IBOS-Sect}

One important security principle adopted by all modern browsers is the
same-origin policy (SOP): it isolates web apps from different origins,
where an origin is represented as a tuple of a protocol, a domain
name, and a port number.  For example, if a user loads a web page from
origin (\texttt{https},\texttt{mybank.com},\texttt{80}) in one tab and
in a separate tab loads a web page from origin
(\texttt{https},\texttt{mybnk.com},\texttt{80}), i.e., a spoofed
domain that omits the `a' in `mybank,' any code originating from the
spoofed domain in the latter tab will not be able to interact with the
former tab.  SOP also ensures that asynchronous JavaScript and XML
(AJAX) network requests from a web app are routed to its origin.
Unfortunately, browser vendors often fail to correctly implement the
SOP policy \cite{Chen07,Schwenk2017}.

The Illinois Browser Operating System (IBOS) is an operating system
and web browser that was designed and modeled in Maude with security
and compatibility as primary goals
\cite{DBLP:conf/osdi/TangMK10,tang-thesis,DBLP:conf/facs2/SasseKMT12}. Unlike
commodity browsers, where security checks are implemented across
millions of lines of code, in IBOS a small trusted computing base of
only 42K code lines is established in the \emph{kernel} through which
all network interaction is filtered. What this means in practice is
that, even if highly complex HTML rendering or JavaScript
interpretation code is compromised, the browser \emph{still} cannot
violate SOP (and several other security properties besides). The
threat model considered here allows for much of the browser code
itself to be compromised while still upholding the security invariants
we describe below.

In the following subsections we survey the IBOS browser and SOP, how
the IBOS browser can be formally specified as a rewriting logic
theory, and how the SOP and other IBOS security properties can be
formally specified as invariants.

\subsection{IBOS System Specification} \label{IBOS-SSpec}

\textbf{IBOS System Design.} The Illinois Browser Operating System is
an operating system and web browser designed to be highly secure as a
browser while maintaining compatibility with modern web apps.  It was
built on top of the micro-kernel L4Ka::Pistachio
\cite{klein2004towards,kolanski2006formalising}, which embraces the
principles of least privilege and privilege separation by separating
operating subsystems into separate daemons that communicate through
the kernel via checked inter-process communication (IPC).  IBOS
directly piggybacks on top of this micro-kernel design by implementing
various browser abstractions, such as the browser chrome and network
connections, as separate components that communicate using L4ka kernel
message passing infrastructure.  Figure \ref{ibos-arch} gives an
overview of the IBOS architecture; as an explanatory aid we highlight
a few key objects:

\begin{figure}
\centering
\includegraphics[scale=.20]{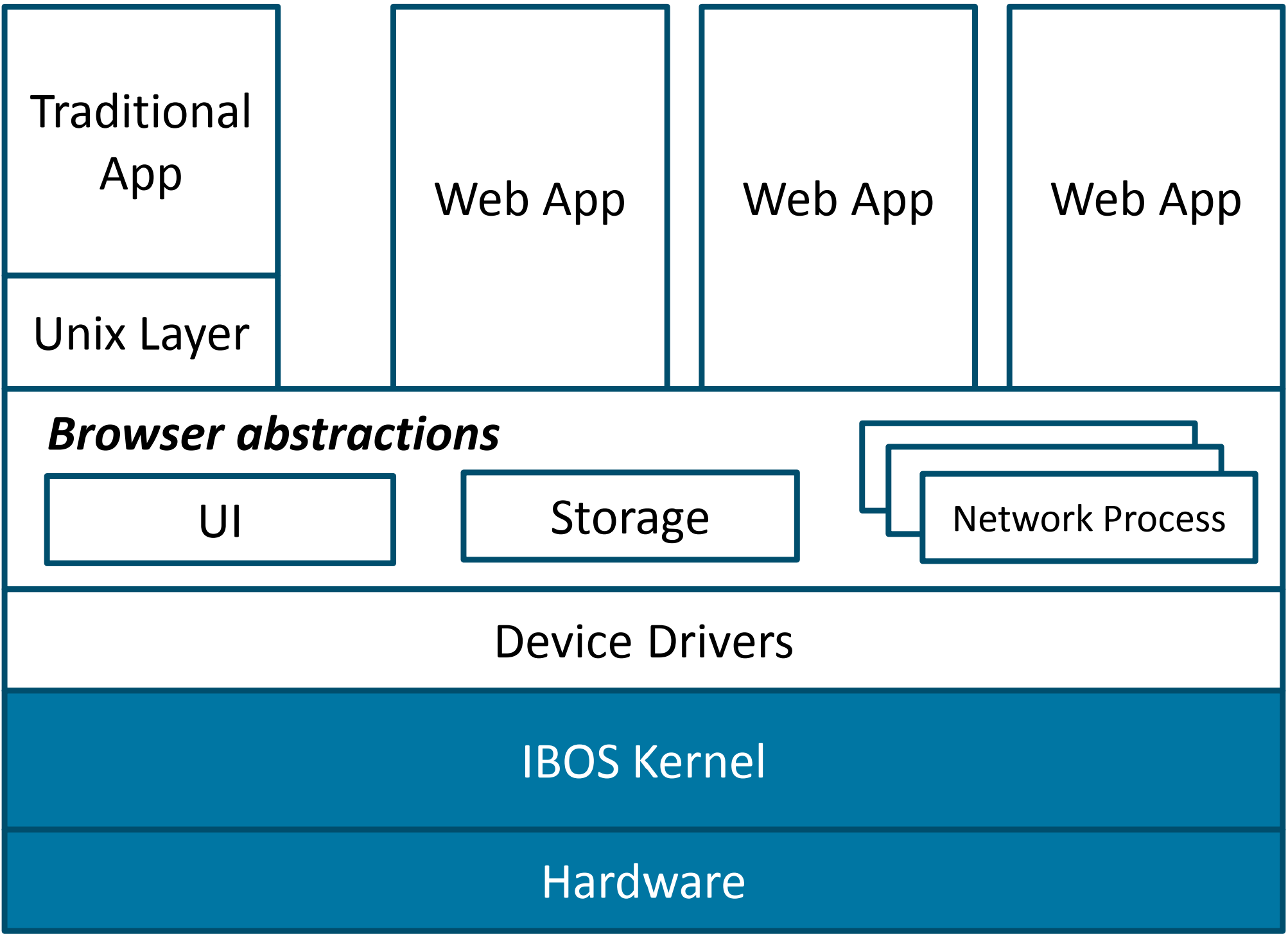}
\caption{IBOS System Architecture \cite{DBLP:conf/facs2/SasseKMT12}.}
\label{ibos-arch}
\end{figure}

\begin{itemize}
\item \textbf{Kernel.} The IBOS kernel is built on top of the
  L4Ka::Pistachio micro-kernel which, as noted previously, can check
  IPC messages against security policies.

\item \textbf{Network Process.} A network process is responsible for
  managing a network connection (e.g., HTTP connections) to a specific
  origin. It understands how to encode and decode TCP datagrams and
  Ethernet frames and can send and receive frames from the network
  interface card (NIC).

\item \textbf{Web Application.} A web application represents a
  specific instance of a web page loaded in a particular browser tab
  (e.g., when a link is clicked or a URL is entered into the address
  bar). Web applications know how to render HTML documents. As per
  SOP, each web page is labeled by its origin.

\item \textbf{Browser UI.} The browser user inferface (UI) minimally
  includes the address bar and the mouse pointer and extends to any
  input mechanism.

\item \textbf{Display.} The display represents the rendered web app
  shown to the user; it is blank when no web app has loaded. For
  security, it cannot modify the UI.
\end{itemize}

\textbf{IBOS System Specification in Maude.}  We present an overview
of the IBOS formal executable specification as a Maude rewrite theory,
which closely follows previous work
\cite{DBLP:conf/facs2/SasseKMT12,sasse-thesis,rocha-thesis}; a more
detailed explanation can be found in Appendix~\ref{sec.appendixB}. We
model IBOS as an object-based rewrite theory (see Section
\ref{prelims}).  We use \textit{italics} to write Maude rewrite rules
and \textit{CamelCase} for variable or sort names.  Some objects of
interest include the singleton objects \textit{kernel}, \textit{ui},
and \textit{display} (in classes \textit{Kernel}, \textit{UI}, and
\textit{Display}, respectively). We also have non-singleton classes
\textit{WebApp} and \textit{NetProc} representing web apps and network
processes in IBOS, respectively.

As an example, let us consider the specification of the
\textit{change-display} rewrite rule shown in
Figure~\ref{change-display-rule}.

\vspace{-25pt}

\begin{figure*}
\begin{align*}
& \; \big\{ \langle \textit{display} \mid \textit{displayContent}(D), \textit{activeTab}(\textit{WA}), \textit{Atts} \rangle \; \langle \textit{WA} \mid \textit{rendered}(D'), \textit{Atts}' \rangle \; \textit{Conf}\ \big\}\\
\rightarrow & \;\big\{ \langle \textit{display} \mid \textit{displayContent}(D'), \textit{activeTab}(\textit{WA}), \textit{Atts} \rangle \; \langle \textit{WA} \mid \textit{rendered}(D'), \textit{Atts}' \rangle \;\textit{Conf}\	\big\} \\
\text{ if } & \; D \neq D'
\end{align*}
\vspace*{-30pt}
\caption{\textit{change-display} rule}
\label{change-display-rule}
\end{figure*}

\noindent This rule involves the \textit{display} object and the
\textit{WebApp} designated as the display's \textit{activeTab}.  In
our model, the \textit{rendered} attribute of a web app represents its
current rendering of the HTML document located at its origin. When the
web app is first created, its \textit{rendered} attribute has the
value \textit{about-blank}, i.e., nothing has yet been rendered. Thus,
this rule essentially states that, at any time, the displayed content
can be replaced by currently rendered HTML document of the active tab,
\emph{only if} it is different from the currently displayed content.

Our IBOS browser specification contains 23 rewrite rules and is about
850 lines of Maude code; it is available at
\url{https://github.com/sskeirik/ibos-case-study}.

\subsection{Specification of the IBOS Security Properties}
\label{ibos-props}

We first describe at a high level the security properties that we will
formally specify and verify. The key property that we verify is the
\emph{same-origin policy} (SOP), but we also specify and verify the
\emph{address bar correctness} (ABC) property.  Our discussion follows
that of \cite{DBLP:conf/facs2/SasseKMT12,sasse-thesis}, based on
invariants $P_1$-$P_{11}$:

\begin{enumerate}[P$_\bgroup1\egroup$]
	\item Network requests from web page instances must be handled by the proper network process.
	\item Ethernet frames from the network interface card (NIC) must be routed to the proper network process.
	\item Ethernet frames from network processes to the NIC must have an IP address and TCP port that matches the origin of the network process.
	\item Data from network processes to web page instances must be from a valid origin.
	\item Network processes for different web page instances must remain isolated.
	\item The browser chrome and web page content displays are isolated.
	\item Only the current tab can access the screen, mouse, and
    keyboard.
	\item All components can only perform their designated functions.
	\item The URL of the active tab is displayed to the user.
	\item The displayed web page content is from the URL shown in the
    address bar.
	\item All configurations are well-formed, i.e., non-duplication of
    \textit{Oid}s and \textit{Att}s.
\end{enumerate}

We define same-origin policy as $\text{SOP}=\bigwedge_{1\leq i \leq 7}
P_i$; address-bar correctness is specified as
$\text{ABC}=P_{10}$. Note that $P_9 \wedge P_{10} \Rightarrow
P_7$. Since $P_5$, $P_6$, and $P_8$ follow directly from the model
design, it is sufficient to prove $\bigwedge_{i\in I} P_i$ for
$I=\{1,2,3,4,9,10\}$. Invariant $P_{11}$ is new to our current
formalization, but is implicitly used in prior work; it forbids absurd
configurations, e.g., having two \emph{kernel}s or a \textit{WebApp}
that has two \textit{URL}s (see Section \ref{prelims}).  Due to its
fundamental relation to how object-based rewrite theories are defined,
we need $P_{11}$ in the proof of \emph{all} other invariants.  As an
example of how these invariant properties can be formalized in our
model as constrained pattern predicates, we show how the address bar
correctness invariant can be specified in our system below:
\begin{multline}
\{ \langle \textit{kernel} \mid \textit{addrBar}(U), \mathit{Atts}
\rangle\\ \langle \textit{display} \mid \textit{displayContent}(U'),
\mathit{Atts}' \rangle \; \mathit{Conf} \} \mid U' \unlhd U \tag{ABC}
\end{multline}
\noindent where $U' \unlhd U \Leftrightarrow_\text{def} (U' \neq
\textit{about-blank} \Rightarrow U = U')$, i.e., the \textit{display}
is either blank \emph{or} its contents' origin matches the address
bar. Note that, for simplicity, in our model, we identify displayed
content with its origin URL. As a second example, consider how to
formalize invariant $P_9$:
\begin{multline}
\big\{ \langle \textit{kernel}  \mid \textit{addrBar}(U), \textit{Atts} \rangle\\
\langle \textit{display} \mid \textit{activeTab}(\textit{WA}), \textit{Atts}' \rangle\\
\langle \textit{WA} \mid \textit{URL}(U''), \textit{Atts}'' \rangle
\;\textit{Conf}\ \big\} \mid U \unlhd U''
\tag{$P_9$}
\end{multline}
\noindent i.e., the address bar must match the URL of the active tab,
unless the address shown is \textit{about-blank}, i.e., nothing is
shown. Finally, $P_{11}$ has a trivial encoding: $\{ \mathit{Conf}
\}\mid \mathit{WF}(\mathit{Conf})$, where $\mathit{WF} : \mathit{Conf}
\rightarrow \mathit{Bool}$ is the well-formedness predicate. Our
specification has 200 lines of Maude code to specify the pattern
predicates used in our invariants and another 900 lines of code
specifying all of the auxiliary functions and predicates.  As stated
in the Introduction, we additionally must prove that: (a) the IBOS
system specification extended by our security property specification
is \emph{suitable} for symbolic reachability analysis; and (b) our ABC
and SOP invariants are at least as strong as the corresponding
invariants in prior work
\cite{DBLP:conf/facs2/SasseKMT12,sasse-thesis} $\text{ABC}_0$ and
$\text{SOP}_0$.  Proof obligation (a) can be met by using techniques
for proving \emph{ground convergence} and \emph{sufficient
  completeness} of conditional equational theories, while proof
obligation (b) can be reduced to proving associated implications,
e.g., $\text{SOP} \Rightarrow \text{SOP}_0$.  We have carried out full
proofs of (a) and (b); due to space limitations, the full details are
available at \cite{skeirik-thesis}.

\section{Proof of IBOS Security Properties} \label{IBOS-Proof}

In this section, due to space limitations, we give a high-level
overview of our proof methodology for verifying the SOP and ABC
properties for IBOS, and show a subproof used in deductively verifying
ABC. Each proof script has roughly 200 lines of code and another 20 to
specify the reachability logic sequent being proved.

\textbf{Modular Proof Methodology.}
In this subsection we survey our modular proof methodology for proving
invariants using reachability logic and comment on how we exploit
modularity in three key ways, i.e.,
we show how we efficiently structured and carried out
our proofs by decomposing them into composable,
independent, and reusable pieces.

Most of the IBOS proof effort was spent strengthening an invariant $I$
into an inductive invariant $I_\mathit{ind}$, where $I$ is either SOP
or ABC.
Typically, $I_\mathit{ind}$ is obtained by iteratively applying the
proof strategy given in Section \ref{prelims}. In each round, assume
candidate $I'$ is inductive and attempt to complete the proof. If,
after applying the \textsc{Step} rule (and possibly some derived
rules), an application of \textsc{Axiom} is impossible, examine the
proof of failed pattern subsumption $\llbracket A \rrbracket \subseteq
\llbracket I' \rrbracket$ in the side-condition of \textsc{Axiom}. If
pattern formula $C$ (possibly using new functions and predicates
defined in theory $\Delta$) can be found that might enable the
subsumption proof to succeed, try again with candidate $I' \wedge
C$. In parallel, our system specification $\mathcal{R}$ is enriched by
extending the underlying \emph{convergent} equational theory
$\mathcal{E}$ to $\mathcal{E}\cup\Delta$ to obtain the enriched
rewrite theory $\mathcal{R}_\Delta$.

The first kind of modularity we exploit is \emph{rule
  modularity}. Recall that any reachability logic proof begins with an
application of the \textsc{Step} rule. Since \textsc{Step} must
consider the result of symbolically rewriting the initial sequent with
all possible rewrite rules, we can equivalently construct our proof on
a ``rule-by-rule'' basis, i.e., if $\mathcal{R}=(\Sigma,E\cup
B,\{l_j\rightarrow r_j \text{ if } \phi_j\}_{j\in J})$, we can
consider $|J|$ separate reachability proofs using respective theories
$\mathcal{R}_j=(\Sigma,E\cup B,\{l_j\rightarrow r_j \text{ if }
\phi_j\})$ for $j\in J$. Thus, we can focus on strengthening invariant
$I'$ for each rule $j\in J$ separately.

Another kind of modularity that we can exploit is \emph{subclass
  modularity}. Because we are reasoning in an object-based rewrite
theory, each rule mentions one or more objects in one or more classes,
and describes how they evolve. Recall from Sect. \ref{prelims} that
subclasses must contain all of the attributes of their superclass, but
may define additional attributes and have additional rewrite rules
which affect them. The upshot of all this is that if we \emph{refine}
our specification by instantiating objects in one class into some
subclass, any invariants proved for the original rules immediately
hold for the same rules in the refined specification. Because of rule
modularity, we need only prove our invariants hold for the newly
defined rules. Even among newly defined rules, all
\emph{non-interfering} rules trivially satisfy any already proved
invariants, where non-interfering rules do not directly or indirectly
influence the state of previously defined attributes.

Lastly, we exploit what we call \emph{structural modularity}. Since
our logic is constructor-based and we assume that $B_\Omega$-matching
is decidable, by pattern matching we can easily specify a set $S$ of
sequents to which we can apply the \emph{same} combination of derived
proof rules. This is based on the intuition that syntactically similar
goals typically can be proved in a similar way. More concretely, given
a set of reachability formulas $S$ and pattern $p\in T_\Omega(X)$, we
can define the subset of formulas $S_p=\{\rrule{(u\mid\varphi)}{B}\in
S\mid \exists\alpha\in[X\rightarrow T_\Omega(X)]\, p\alpha=_{B_\Omega
}u\}$. Although the simplified example below does not illustrate
structural modularity, we have heavily exploited this principle in our
formal verification of IBOS.

\textbf{Address Bar Correctness Proof Example.}  Here we show a
snippet of the ABC invariant verification, namely, we prove that the
invariant holds for the \textit{change-display} rule by exploiting
rule modularity as noted above. As mentioned in our description of
invariant $P_{11}$, well-formedness is required for all
invariants. Therefore, we begin with $\text{ABC} \wedge P_{11}$ as our
candidate inductive invariant, which, as mentioned in
Sect. \ref{prelims}, normalizes by disjoint $B_\Omega$-unification to
the invariant:

\noindent \begin{multline*}
\big\{ \langle \textit{kernel}  \mid \textit{addrBar}(U), \textit{Atts} \rangle\\
\langle \textit{display} \mid \textit{displayContent}(U'), \textit{Atts}' \rangle \; \textit{Conf}\ \big\} \mid U' \unlhd U \wedge \mathit{WF}(\ldots)
\end{multline*}

\noindent where for brevity $\ldots$ expands to the entire term of
sort \textit{Conf} wrapped inside operator $\{\_\}$, i.e., the entire
configuration is \emph{well-formed}. Recall the definition of the
\textit{change-display} rule in Section \ref{IBOS-SSpec}.  We can see
that our invariant only mentions the \textit{kernel} and
\textit{display} processes, whereas in rule \textit{change-display}
the value of \textit{displayContent} depends on the \textit{rendered}
attribute of a \textit{WebApp}, i.e., the one selected as the
\textit{activeTab}.  Clearly, our invariant seems too weak. How can we
strengthen it? The reader may recall $P_9$, which states ``the URL of
the active tab is displayed to the user.''  Thus, by further disjoint
$B_\Omega$-unification, the strengthened invariant $\text{ABC}\wedge
P_{11} \wedge P_9$ normalizes to:

\noindent \begin{multline*}
\big\{ \langle \textit{kernel}  \mid \textit{addrBar}(U), \textit{Atts} \rangle \; \langle \textit{display} \mid \textit{displayContent}(U'), \textit{activeTab}(\textit{WA}), \textit{Atts}' \rangle\\
\langle \textit{WA} \mid \textit{URL}(U''), \textit{Atts}'' \rangle \;
\textit{Conf}\ \big\} \mid U' \unlhd U \wedge U \unlhd U'' \wedge \mathit{WF}(\ldots)
\end{multline*}
This new invariant is closer to what we need, since the pattern now
mentions the particular web app we want. Unfortunately, since our
invariant still knows nothing about the \textit{rendered} attribute,
at least one further strengthening is needed.

At this point, we can enrich our theory with a new predicate stating
that the \textit{rendered} and \textit{URL} attributes of any
\textit{WebApp} always agree\footnote{Note that, in a very real
  sense, this requirement is at the heart of the SOP, since it means
  that any \textit{WebApp} has indeed obtained content from its
  claimed origin.}.  Let us declare it as $\textit{R} : \textit{Conf}
\rightarrow \textit{Bool}$.  We can define it inductively over
configurations by:

\begin{align}
\mathit{R}(\langle \textit{WA} \mid \textit{rendered}(U), \textit{URL}(U'), \textit{Atts} \rangle \; \mathit{Conf}) &= U \unlhd U' \wedge \mathit{R}(\mathit{Conf})\tag{$R_1$}\\
\mathit{R}(\langle \textit{P} \mid \textit{Atts} \rangle \; \mathit{Conf}) &= \mathit{R}(\mathit{Conf}) \textit{ if } \neg\textit{WA}(P)\tag{$R_2$}\\
\mathit{R}(\mathit{none}) &= \top\tag{$R_3$}
\end{align}

\noindent where $\textit{WA} : \textit{Oid} \rightarrow \textit{Bool}$
ambiguously denotes a predicate that holds iff an \textit{Oid} refers
to a web app. Intuitively, it says that whatever a \textit{WebApp} has
\textit{rendered} is either blank or has been
 loaded from its \textit{URL}. The strengthened invariant becomes:

\noindent \begin{multline*}
\big\{ \langle \textit{kernel}  \mid \textit{addrBar}(U), \textit{Atts} \rangle \; \langle \textit{display} \mid \textit{displayContent}(U'), \textit{activeTab}(\textit{WA}), \textit{Atts}' \rangle\\
\langle \textit{WA} \mid \textit{URL}(U''), \textit{Atts}'' \rangle \;
\textit{Conf}\ \big\} \mid\\
U' \unlhd U \wedge U \unlhd U'' \wedge \mathit{WF}(\ldots) \wedge R(\ldots)
\end{multline*}

\noindent where, as before, the $\ldots$ represents the entire term of
sort \textit{Conf} enclosed in $\{\_\}$. Now, all of the required
relationships between variables in the rewrite rule seem to be
accounted for. Uneventfully, with the strengthened invariant $ABC
\wedge P_{11} \wedge P_{9} \wedge R$, the subproof for the
\emph{change-display} rule now succeeds.

\section{Related Work and Conclusions} \label{related-concl}

{\bf Related Work on IBOS Verification}.  In this paper, we have
presented the first full deductive verification of SOP and ABC for
IBOS. Note that in \cite{DBLP:conf/facs2/SasseKMT12,sasse-thesis}, SOP
and ABC were also verified. Their approach consisted of a hand-written
proof that any counter-example must appear on some trace of length $n$
plus bounded model checking showing that such counterexamples are
unreachable on all traces of length $n$.  In \cite{rocha-thesis}, an
attempt was made to deductively verify these same invariants via the
Maude invariant analyzer. Though a few basic invariants were proved,
due to thousands of generated proof obligations, none of the
properties listed in Sect. \ref{ibos-props} were verified.  Compared
to previous work, this paper presents the first full deductive
verification of IBOS security properties.

\vspace{1ex}

\noindent {\bf Related Work on Browser Security}.  In terms of
computer technology, the same-origin policy is quite old: it was first
referenced in Netscape Navigator 2 released in 1996 \cite{JSGuideV12}.
As \cite{Schwenk2017} points out, different browser vendors have
different notions of SOP.
Here, we situate IBOS and our work into this larger context.
Many papers have been written on policies for enforcing SOP with
regards to frames \cite{barth2009securing}, third-party scripts
\cite{jackson2008beware,karlof2007dynamic}, cached content
\cite{jackson2006protecting}, CSS
\cite{Huang:2010:PBC:1866307.1866376}, and mobile OSes
\cite{Wang:2013:UOC:2508859.2516727}.
Typically, these discussions assume that browser code is working as
intended and then show existing policies are insufficient. Instead,
IBOS attacks the problem taking a different tack: even if the browser
itself is compromised, can SOP still be ensured? What the IBOS
verification demonstrates is that ---by using a minimal trusted
computing base in the kernel, implementing separate web frames as
separate processes, and requiring all IPC to be kernel-checked--- one
can in fact enforce the standard SOP notions, \emph{even if} the
complex browser code for rendering HTML or executing JavaScript is
\emph{compromised}.
Although our model does not treat JavaScript, HTML, or cookies,
explicitly, since it models system calls which are used for process
creation, network access, and inter-process communication, code
execution and resource references can be abstracted away into the
communication primitives they ultimately cause to be invoked, allowing
us to perform strong verification in a high-level fashion.
\cite{bugliesi2017formal} surveys many promising lines of research in
the formal methods web security landscape. Prior work on formal and
declarative models of web browsers includes \cite{bauer2015run} as
well as the \emph{executable} models
\cite{bohannon2010featherweight,bohannon2012foundations}. Our work
complements the Quark browser design and implementation of
\cite{jang2012establishing}: Quark, like IBOS, has a small trusted
kernel (specified in Coq). In addition to proving tab non-interference
and address bar correctness theorems, the authors
use Coq code extraction to produce a verified, functional
browser. Unlike Quark, whose TCB includes the entire Linux kernel
and Coq code extraction tools, the TCB
of the IBOS browser consists of only 42K lines of C/C++ code.

\vspace{1ex}

\noindent {\bf Related Work on Reachability Logic}.  Our work on
\emph{constructor-based} reachability logic
\cite{DBLP:conf/lopstr/SkeirikSM17,DBLP:journals/fuin/SkeirikSM20} 
builds upon previous work on
reachability
logic~\cite{DBLP:conf/fm/RosuS12,DBLP:conf/oopsla/RosuS12,DBLP:conf/rta/StefanescuCMMSR14,DBLP:conf/oopsla/StefanescuPYLR16}
as a language-generic approach to program verification, parametric on
the operational semantics of a programming language.  Our work extends
in a non-trivial manner reachability logic from a
programming-language-generic logic of programs to a
rewrite-theory-generic logic to reason about \emph{both} distributed
system designs and programs based on rewriting logic semantics.  Our
work in
\cite{DBLP:conf/lopstr/SkeirikSM17,DBLP:journals/fuin/SkeirikSM20} 
was also inspired by the
work in \cite{DBLP:conf/birthday/LucanuRAN15}, which for the first
time considered reachability logic for rewrite theories, but went
beyond \cite{DBLP:conf/birthday/LucanuRAN15} in several ways,
including more expressive input theories and state predicates, and a
simple inference system as opposed to an algorithm.  Also related to
our work in
\cite{DBLP:conf/lopstr/SkeirikSM17,DBLP:journals/fuin/SkeirikSM20}, 
but focusing on
\emph{coinductive reasoning}, we have the recent work in
\cite{moore-thesis,DBLP:journals/jsc/LucanuRA17,DBLP:conf/cade/CiobacaL18},
of which, in spite of various substantial differences, the closest to
our work regarding the models assumed, the kinds of reachability
properties proved, and the state predicates and inference systems
proposed is the work in \cite{DBLP:conf/cade/CiobacaL18}.

\vspace{1ex}

\noindent\textbf{Conclusion and Future Work}.  We have presented a
full deductive proof of the SOP and ABC properties of the IBOS browser
design, as well as a prover and a modular reachability logic
verification methodology making proofs scalable to substantial proof
efforts like that of IBOS.  Besides offering a case study that can
help other distributed system verification efforts, this work should
also be seen as a useful step towards incorporating the IBOS design
ideas into future fully verified browsers.  The web is alive and
evolving; these evolution necessitates that formal approaches evolve
as well.  Looking towards the future of IBOS, two goals stand out: (i)
extending the design of IBOS to handle some recent extensions of the
SOP, e.g., cross-origin resource sharing (CORS) to analyze potential
cross-site scripting (XSS) and cross-site request forgery attacks
(XSRF)~\cite{gollmann-sop-2011}, and to check for incompatible content
security policies (CSP)~\cite{bielova-csp-2017} in relation to SOP; by
exploiting subclass and rule modularity, the verification of an IBOS
extension with such new functionality could reuse most of the current
IBOS proofs, since extra proofs would only be needed for the new,
functionality-adding rules; and (ii) exploiting the intrinsic
concurrency of Maude rewrite theories to transform them into
correct-by-construction, deployable Maude-based distributed system
implementations, closing the gap between verified designs and correct
implementations.
Our work on IBOS takes one more step towards demonstrating that a
formally secure web is possible in a connected world where security is
needed more than ever before.

\bibliographystyle{splncs04}
\bibliography{ref}

\appendix

\section{Reachability Logic Derived Proof Rules} \label{der-pr-rules}

For completeness, we document the three derived rules in our proof system in Table \ref{DerivedRulesOverview}, provide some intuitive explanations of how they are used, and finally make a few general comments about how they are automated.

{\captionsetup{justification=raggedright,labelsep=none,labelformat=empty,singlelinecheck=off}
	\begin{table*}
		\raggedright
		\captionof{table}{\hspace{29pt} \textbf{\ul{Table 2: Overivew of Derived Proof Rules}}}\label{DerivedRulesOverview}

		\vspace{5pt}
		\hspace{29pt} \textit{Assumption:} $\mathcal{R}=(\Sigma,E\!\cup\! B,R)$ is suff. comp. w.r.t. $(\Omega,B_\Omega)$
		\vspace{5pt}

		\centering

		\begin{tabular}{@{}lcl@{}}
			\toprule
			Name & Rule & Condition\\
			\midrule
			\textsc{Case}$^\text{*}$

			& $\logicrule{
				\bigwedge_{a\in M} \ssequent{\cA}{\!\cC}{\rrule{\!A[x/a]}{B[x/a]}}
			}{
				\sequent{\cA}{\cC}{\!A}{B}
			}$\vspace*{10pt}

			& $x\!:\!s\in\!V \wedge \mathit{cover}(M,s)$\\

			\textsc{Split}$^\dagger$

			& $\logicrule{
				\bigwedge_{\phi\in \Phi} \ssequent{\cA}{\!\cC}{\!\rrule{(u\mid \varphi \wedge \phi)}{B}}
			}{
				\sequent{\cA}{\cC}{\!(u\mid\varphi)}{B}
			}$\vspace*{10pt}

			& $(\bigvee \Phi = \top) \wedge \mathit{away}(\Phi,A,B)$ \\

			\textsc{Subst.}$^\ddagger$\

			& $\logicrule{
				\bigwedge_{\alpha\in\mathit{Unif}_{\!\!E}\!(\psi)} \ssequent{\cA}{\!\cC}{\!\rrule{(u\mid \rho)\alpha}{B\alpha}}
			}{
				\sequent{\cA}{\cC}{\!(u\mid\varphi)}{B}
			}$\ \ \vspace*{3pt}

			& $\varphi = \psi \wedge \rho$ \\
			\bottomrule
		\end{tabular}\phantom{a}\smallskip\\

		\raggedright
		\hspace{29pt} $^\text{*}\ \mathit{cover}(M,s) \equiv M\subseteq  T_\Omega(Y)_s \wedge Y \cap \vars{A,B} = \emptyset \wedge
		\llbracket M \rrbracket = T_{\Omega,s} \wedge | M | \text{ finite }$\\
		\hspace{29pt} $^\dagger\ \mathit{away}(\Phi,A,B) \equiv \vars{\Phi}\cap(\vars{B}/\vars{A}) = \emptyset$.\\
		\hspace{29pt} $^\ddagger\ \text{Usable only when } \mathit{Unif}_{\!E}(\psi)$ is decidable.
	\end{table*}
}

In some sense, the \textsc{Case} and \textsc{Split} rules are
performing the exact same operation. In the former, we are considering
how to decompose a variable into multiple patterns which entirely
\textit{cover} the sort of the given variable; in the later, we are
considering how we can decompose the true formula $\top$ into a set of
quantifier-free formulas whose disjunction is equivalent to $\top$. In
either case, what we want is to make use of the extra information
available after case analysis or formula splitting to help us
discharge pattern subsumption proofs in the side-condition of
\textsc{Axiom}. These rules are both essential for our proofs of both
SOP and ABC, since they allow us to infer additional information in a
sound way that is implied by application of the system rewrite rules
but not directly stated.
The \textsc{Substitution} rule has a slightly different flavor from
\textsc{Case} or \textsc{Split}; essentially, it encodes a commonly
applicable sequent simplification technique when a fragment of the
sequent constraint is a \emph{unifiable} conjunction of equalities.
Its usefulness primarily comes from the fact that such unifications
often \emph{fail}, which can lead to a huge reduction in the number of
goals to be proved.

As for automation, these rules present no special difficulties.
\textsc{Substitution} is implementable directly by Maude's built-in
unification modulo axioms and variant unification algorithms. The
\textsc{Case} and \textsc{Split} rules are also easy to apply.
Regarding the checking of their
side conditions,  checking in \textsc{Case} that  $M$ is a cover set for
sort $s$ is easy to automate  by
tree automata modulo axioms $B$ techniques
supported by Maude's Sufficient Completeness Checker
\cite{hendrix-meseguer-ohsaki-ijcar06,hendrix-thesis}.
In the case of
\textsc{Split}, although proving that an \emph{arbitrary} disjunction
$\bigvee \Phi$
of formulas is inductively equivalent
to $\top$ in the initial algebra $T_{\Sigma/E \cup B}$
 is in general undecidable, in practice $\bigvee \Phi$
is \emph{not arbitrary at all}, since  $\bigvee \Phi$ typically  has
the form $p= \mathit{true} \vee p= \mathit{false}$, where $p$
is a Boolean expression, or the form  $F \vee \neg F$, for $F$ a QF-formula;
so such checks are trivial in practice.

\section{IBOS System Specification in Maude}
\label{sec.appendixB}

To ground our discussion we recap here some details of the IBOS specification.
The Maude rewrite theory specifying the IBOS system design in total consists of approximately 850 lines of code and defines 23 rewrite rules.
The main pattern
predicate defining the SOP is about 180 lines of code while the main pattern
 predicate defining ABC is about 20 lines of code. Their supporting predicate definitions used to define the inductive invariant is about 900 lines of code.
The reachability formulas specifying the inductive invariants for the IBOS SOP and ABC security properties in our Maude reachability logic tool notation are roughly 20 lines each.
The proof scripts which verify that the respective inductive invariants hold for each proof rule excluding comments and boilerplate text are about 200 lines each.
In total, there are approximately 2500 lines of code.

As a further aid to the reader and a complement to the graphical
overview of IBOS in Figure \ref{ibos-arch}, we rewrite this graphical
figure using our formal specification.
Said another way, we provide in Figure \ref{IBOSRepState} a representative state (e.g., a ground term) of the transition system $(C_{\Sigma/E\cup B,\mathit{State}},\rightarrow_\mathcal{R})$ where $\mathcal{R}$ is the Maude rewrite theory specifying IBOS. To improve readability, we write each object attribute on a separate line.

Let us make a few high-level remarks about the figure.
In our specification, urls are encoded as numbers wrapped by constructor $\mathit{url(\ldots)}$.
In IBOS, to enforce SOP as well as other security policies, both browser frames and network connections must be tracked.
Each browser frame is represented by an object of class \textit{WebApp}, while each network connection is represented by an object of class \textit{NetProc}.
The \textit{kernel} manages process state by internal metadata tables \textit{webAppInfo} where each web app is tagged by its origin and \textit{netProcInfo} where each network connection is tagged by its origin web app and destination server.
When a new web app is created, the kernel automatically creates a corresponding network process between the webapp and its origin server.
Since the \textit{kernel} is responsible for creating network connections, it records the next fresh network process identifier in \textit{nextNetProc}.

The \textit{kernel}'s \textit{secPolicy} attribute stores its security policy. Each policy consists of a sender, a receiver, and a message type. Any message not explicitly allowed by the policy is dropped.
In the policy, the special \textit{Oid}s \textit{network} and \textit{webapp} represent a policy allowing any network process or webapp to send a particular kind of message to its corresponding webapp or network process, respectively.
In the figure, the \textit{kernel} is preparing to forward a message from $\mathit{webapp(0)}$ to its corresponding network process $\mathit{network(0)}$ asking to fetch an item from the web app's origin $\mathit{url(15)}$.

Aside from the \textit{kernel}, the system has a few other distinguished objects.
The \textit{display} object tracks the content currently shown on the screen in \textit{displayContent}; to do that, it should know which web app is the \textit{activeTab}. The \textit{ui} (user interface) contains the list of commands given by the user during some usage session in its \textit{toKernel} attribute. The \textit{webappmgr} (web app manager) is responsible for spawning new web apps; in our model, it just records the next fresh web app identifier in attribute \textit{nextWebApp}. Finally, the \textit{nic} (network interface card) has two attributes for ingoing and outgoing data. In our model, we identify urls and their loaded content. To model network latency, outgoing messages in \textit{nic-out} are queued up in \textit{nic-in} in a random order.

Of course, we also have objects representing web apps and network connections.
In the figure, $\mathit{webapp(0)}$'s \textit{rendered} content is blank. However, it is currently \textit{loading} its content from its origin \textit{url(15)}; its request to fetch data from its corresponding network process is currently being handled by the kernel. Currently, its \textit{toKernel} and \textit{fromKernel} IPC message queues are blank because it is not performing any other communication at this time.
There is also a corresponding network process \textit{network(0)}; it has two direct-memory access (DMA) buffers \textit{mem-in} and \textit{mem-out} that are used as I/O channels between itself and the network card driver. Like web apps, network processes also have two IPC message queues. Finally, a network process also stores which web app its received data should be sent back to.

\begin{figure}
	\mathchardef\mhyphen="2D
	\centering
	\hspace*{-20pt}
	\begin{minipage}{0.5\textwidth}
		$\mathit{< kernel  \mid addrBar(url(15)),}\\
		\hspace*{40pt}\mathit{handling(msg(webapp(0),network(0),FETCH\mhyphen URL,url(15)))}\\
		\hspace*{40pt}\mathit{nextNetProc(1),}\\
		\hspace*{40pt}\mathit{webAppInfo(pi(webapp(0),url(15))),}\\
		\hspace*{40pt}\mathit{netProcInfo(pi(network(0),url(15),url(15))),}\\
		\hspace*{40pt}\mathit{secPolicy(policy(webapp,network,FETCH\mhyphen URL),}\\
		\hspace*{50pt}\mathit{policy(network,webapp,RETURN\mhyphen URL),}\\
		\hspace*{50pt}\mathit{policy(ui,webapp,NEW\mhyphen URL),}\\
		\hspace*{50pt}\mathit{policy(ui,webapp,SWITCH\mhyphen TAB)) >}\\
		\mathit{< display \mid displayContent(about\mhyphen blank),}\\
		\hspace*{43pt}\mathit{activeTab(webapp(0)) >}\\
		\mathit{< ui \mid toKernel(msg(ui,webapp,NEW\mhyphen URL,url(25))) >}\\
		\mathit{< webappmgr \mid nextWebApp(1) >}\\
		\mathit{< nic \mid nic\mhyphen in(emtpy),}\\
		\hspace*{28pt}\mathit{nic\mhyphen out(empty) > }\\
		\mathit{< webapp(0) \mid URL(url(15)),}\\
		\hspace*{58pt}\mathit{rendered(about\mhyphen blank),}\\
		\hspace*{58pt}\mathit{loading(true),}\\
		\hspace*{58pt}\mathit{toKernel(empty),}\\
		\hspace*{58pt}\mathit{fromKernel(empty) >}\\
		\mathit{< network(0) \mid mem\mhyphen in(empty),}\\
		\hspace*{58pt}\mathit{mem\mhyphen out(empty),}\\
		\hspace*{58pt}\mathit{returnTo(webapp(0)),}\\
		\hspace*{58pt}\mathit{toKernel(empty),}\\
		\hspace*{58pt}\mathit{fromKernel(empty) >}$
	\end{minipage}
	\caption{Representative IBOS System State}
	\label{IBOSRepState}
\end{figure}

\end{document}